\documentclass[manuscript,screen,review=false]{acmart}
\AtBeginDocument{%
  \providecommand\BibTeX{{%
    \normalfont B\kern-0.5em{\scshape i\kern-0.25em b}\kern-0.8em\TeX}}}

\setcopyright{acmcopyright}
\copyrightyear{2023}
\acmYear{2023}
\acmDOI{XXXXXXX.XXXXXXX}

\acmConference[SSIP '23]{ACM Conference}{October,
  2023}{Nanjing, China}
\acmBooktitle{Woodstock '18: ACM Symposium on Neural Gaze Detection,
 June 03--05, 2018, Woodstock, NY} 
\acmPrice{15.00}
\acmISBN{978-1-4503-XXXX-X/18/06}

\acmSubmissionID{NA0003}

\usepackage{amsmath}
\DeclareMathOperator{\dotprod}{\cdot}

\begin{document}
\settopmatter{printacmref=false} 

\title{Attention-based Efficient Classification for 3D MRI Image of Alzheimer’s Disease}

\author{Yihao Lin*}\footnote{*These authors contributed to the work equllly and should be regarded as co-first authors.}
\affiliation{%
  \institution{Faculty of Computing, Harbin Institute of Technology}
  \city{Harbin}
  \country{China}}
\email{lyh01hao@gmail.com}
 
\author{Ximeng Li*}
\affiliation{%
  \institution{Harbin No.3 High School}
  \city{Harbin}
  \country{China}}
\email{1454979015@qq.com}
 
\author{Yan Zhang}
\affiliation{%
  \institution{School of Life Science and Technology, Harbin Institute of Technology}
  \city{Harbin}
  \country{China}}
\email{zhangtyo@hit.edu.cn}

\author{Jinshan Tang}
\affiliation{%
  \institution{College of Public Health, George Mason University}
  \city{Fairfax}
  \country{America}}
\email{jtang25@gmu.edu}


\begin{abstract}
  Early diagnosis of Alzheimer Diagnostics (AD) is a challenging task due to its subtle and complex clinical symptoms. Deep learning-assisted medical diagnosis using image recognition techniques has become an important research topic in this field. The features have to accurately capture main variations of anatomical brain structures. However, time-consuming is expensive for feature extraction by deep learning training. This study proposes a novel Alzheimer's disease detection model based on Convolutional Neural Networks. The model utilizes a pre-trained ResNet network as the backbone, incorporating post-fusion algorithm for 3D medical images and attention mechanisms. The experimental results indicate that the employed 2D fusion algorithm effectively improves the model's training expense. And the introduced attention mechanism accurately weights important regions in images, further enhancing the model's diagnostic accuracy.
\end{abstract}

\maketitle

\section{Introduction}
Alzheimer's disease (AD) is a progressive neurodegenerative disorder and the most prevalent cause of dementia in later life. Early and accurate diagnosis of AD is crucial, as the disease dramatically reduces brain volume over time, affecting numerous cognitive functions. Magnetic resonance imaging (MRI) has become a widely employed neuroimaging technology for AD diagnosis. In recent years, machine learning techniques applied to MRI data have demonstrated promising results in medical imaging analysis and diagnostic tasks\cite{hazarika2021improved} \cite{afzal2019data} \cite{prakash2019comparative}\cite{dwivedi2022multimodal}.\par
Some learning-based Alzheimer's disease (AD) studies utilize regions of interest (ROIs) extracted from the original MRI as input for neural network models \cite{ortiz2014automatic}\cite{rondina2018selecting}\cite{duraisamy2019alzheimer}. Methods for selecting ROIs in brain MRI for diagnostic purposes have been presented, employing learning and vector quantization techniques. Compared to whole-brain learning methods, the ROI-learning approach has yielded improved classification accuracies. However, the manual selection of ROI inputs can be time-consuming\cite{XiaomingLiu2023b}\cite{XiaomingLiu2023c}. \par 
Some researchers have used whole images as input for CNN models to eliminate the manual feature extraction step for ROIs in brain MRI images. 3D CNN models have gained attention for diagnosing AD patients using MRI images, as MRI technology produces 3D images comprising millions of voxels that correspond to physical locations in patients' brains \cite{cheng2017classification}\cite{esmaeilzadeh2018end}\cite{albright2019forecasting}\cite{pan2019early}\cite{xing2020dynamic}\cite{Xing2021.05.24.21257554}. Korolev et al. \cite{korolev2017residual} proposed two 3D CNN architectures based on VGGNet and ResNet for AD classification, demonstrating that manual feature extraction is not necessary for brain MRI image classification. Cheng et al. \cite{cheng2017classification} suggested using multiple 3D CNN models trained on MRI images for AD classification within an ensemble learning strategy. Kompanek et al.\cite{kompanek2019volumetrie}used pre-processed MRI 3D image data and a 3D Convolutional Neural Network-based method for binary classification of Alzheimer's disease. However, these 3D CNN models are computationally expensive and time-consuming to train due to the high dimensionality of the input data.\par
For reducing the dimensionality of 3D-MRI volumes, Xing et al. \cite{xing2020dynamic} proposed applying approximate rank pooling to convert a 3D MRI volume into a 2D image over the height dimension. Nonetheless, approximate rank pooling cannot achieve the same accuracy as the ROI learning method, as critical image components are not effectively extracted, and irrelevant information is not disregarded.\par
To balance timing consumption and performance for classification, we present an AD detection model for 3D MRI images to efficiently fusion features with image information. The main contributions of this paper are as follows:

\begin{itemize}
    \item Post-fusion method is given for feature extraction to reduce computing expense and filter low confidence information.

    \item An attention module is presented to increase classification accuracy by comprising spatial and channel attention, focusing more on the critical parts of images while ignoring irrelevant information.

    \item Comparison experiments are conducted with the original image model, and the results showed that our approach achieved better recognition performance with less computing expense.
\end{itemize}

\section{Related Work}

Alzheimer's disease clinical diagnosis requires a comprehensive analysis that combines patient MRI images and reports, supplemented by additional data such as assessment scales. Research on Alzheimer's disease classification primarily focuses on the diagnostic classification of single-modal brain imaging data, emphasizing classification and evaluation through the examination of hippocampal atrophy in the brain.

Deep learning, a subfield of machine learning, gained significant attention after the groundbreaking performance of the AlexNet model, created by Alex's team, in the 2012 ImageNet Large Scale Visual Recognition Challenge. Deep learning model outperformed traditional machine learning techniques 
\cite{Tang2011SVM}\cite{Xu2008ANN}
by a substantial margin in terms of classification accuracy. Since then, deep learning has been extensively investigated and has found various applications in fields such as speech recognition, image recognition, autonomous driving, natural language processing, and so on\cite {Hua2020mutlimedia}\cite{LiHangmutlimedia}. In medical imaging, commonly used models include LeNet, AlexNet, VGG, ResNet, and 3D-CNN.

Hazarika et al.\cite{hazarika2021improved} combined a single maximum pooling layer in LeNet with an additional minimum pooling layer, retaining low-intensity pixel values overlooked by the maximum pooling layer for Alzheimer's disease classification diagnosis. The modified LeNet model maintained its simplicity, effectiveness, and rapid computational performance while demonstrating promising results. Afzal et al.\cite{afzal2019data} employed a pre-trained AlexNet model in their research, selecting 199 sample data from the OASIS Alzheimer's disease research dataset for binary classification between AD and NC. Their work achieved high classification accuracy.


Several studies have focused on the application of deep learning algorithms for the diagnosis and classification of Alzheimer's disease (AD) using structural magnetic resonance imaging (sMRI) data. S. Wang et al.\cite{wang2017automatic} employed a transfer learning approach with pre-training on the OASIS and LIDC datasets, both of which contain sMRI data. Their method achieved a classification accuracy of 90.6\% in differentiating between subjects with mild cognitive impairment (MCI) and normal controls (NC).

In recent work, Tufail et al.\cite{tufail2020binary} proposed the use of multiple deep 2D neural networks for binary AD/NC classification. They introduced two architectures, based on the InceptionV3 and Xception models, respectively, utilizing transfer learning with weights pre-trained on the Imagenet LSVRC dataset. Their approach demonstrated promising results and achieved good performance in AD classification.

Another study by Ghazal et al.\cite{ghazal2022alzheimer} utilized the pre-trained AlexNet for the classification task. The CNN was retrained and validated on a separate validation dataset, achieving an accuracy of 91.7\% for multi-class problems after 40 epochs. Notably, their proposed system model (ADDTLA) eliminated the need for hand-crafted features and exhibited efficiency in handling small image datasets.

\section{Proposed Classification Method}
The problem is to transform 3D MRI images into a 2D images while preserving the relevant features of the original image as much as possible. The input is an MRI 3D image. The image can be sliced at different depths to generate a set of two-dimensional images with a sequential sequence, used as the input to the model. In this part, we will introduce the framework of the proposed model, which consists of a CNN feature extractor, an image fusion module, an attention mechanism module, and a fully connected classifier. ResNet model is used as a pre-trained CNN model for feature extractor. The structure is shown in the Fig 1.

\subsection{Post-fusion feature extraction}
Post-fusion is used to take each two-dimensional slice of the original three-dimensional image as the input of the convolutional neural network. Further fusion is performed after the convolutional neural network outputs the results to classify the original three-dimensional image. A rank pooling method merges multiple 2D slices into a single 2D image. This method was initially designed for video recognition. Analogous to 3D images, the time dimension in the video can be considered as the Z axis in the 3D image. 

\begin{figure}[htp]
    \centering
    \includegraphics[width=6cm]{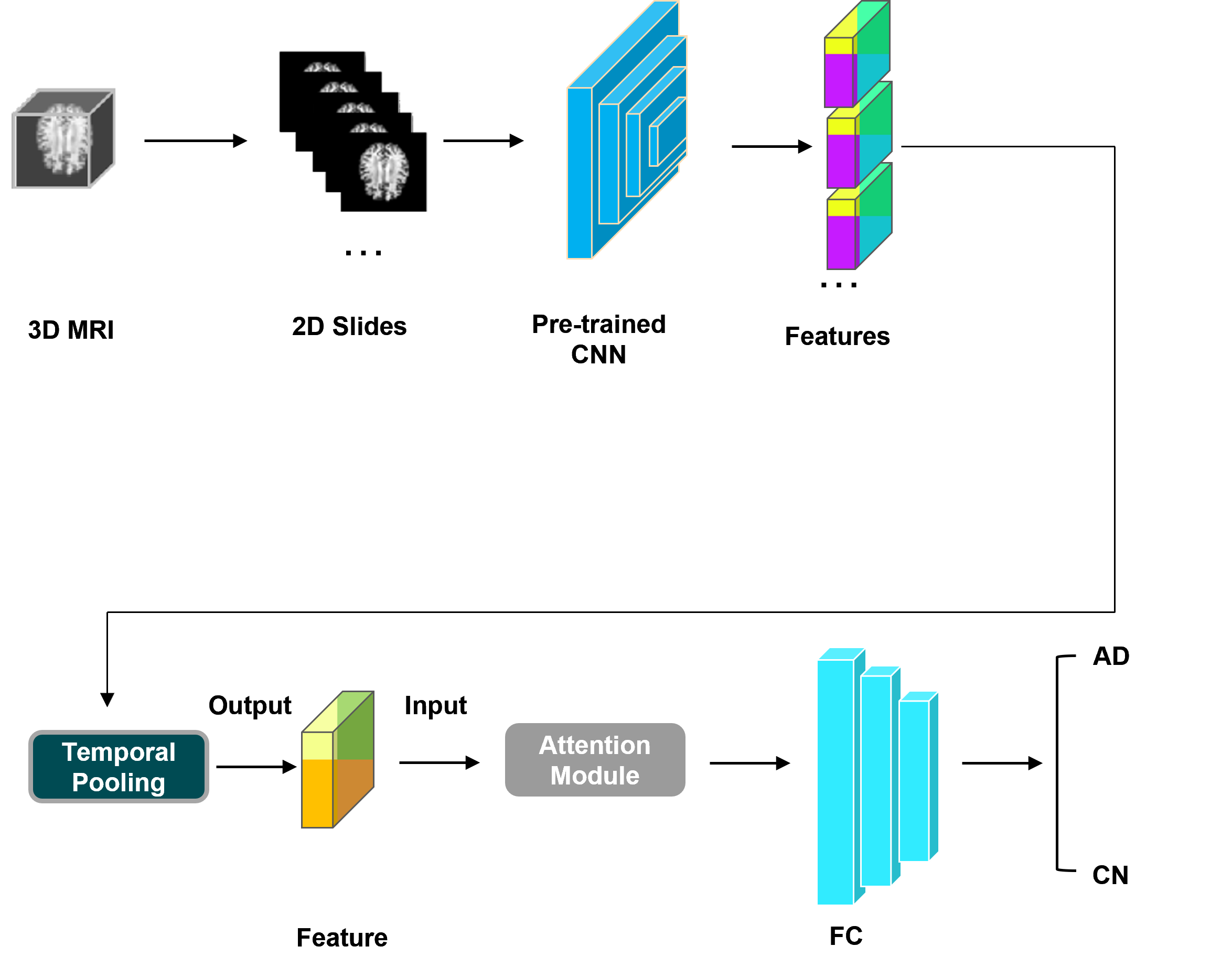}
    \caption{Post-fusion strategy}
    \label{fig:galaxy}
\end{figure}
This method uses a pooling method to represent 3D image.  $\varphi\left(I_t\right)$ is a feature of slice $I_t$. A smooth operation is performed on these slices, such as a mean vector that changes over time: 
\begin{equation}
    v_t=\frac{1}{t}\sum_{\tau=1}^{t}\varphi\left(I_\tau\right)
\end{equation} where $v_t$ is the average depth feature of the frame from the original depth to depth t. The new sequence $V=\left[v_1,v_2,\ldots,v_n\right]$ obtained by the smooth operation is the same size as the original sequence of 2D slices. It can be considered as a new feature vector of the original 2D slices sequence. The goal of this method is to encode the evolution of the appearances of slices in different depths. We define this evolution as dynamic $D$ which reflects how the vector changes for inputs from depth t to t+1. And the dynamic information of the video encoded using a linear sorting function:
\begin{equation}
    \Psi_u=\Psi\left(V;u\right)
\end{equation} where u is the parameter of the sorting function to be learned. This parameter represents the entire video to achieve the goal of dimension reduction for the 3D image. In this function, $\Psi$ approximates $D$. For each slice, a score obtained by the sorting function is set to $S(t,u) = u^T \dotprod v_t$. t is a specific depth. The deeper the depth, the greater the corresponding score, i.e. $ q > t \rightarrow S(q|u) > S(t|u)$. It is the constraints of the ranking problem. The target of ranking pooling is:
\begin{equation}
    argmin_u\frac{1}{2}||u||^2+\lambda \sum_{q>t} \epsilon_{qt}\\
\end{equation}

\begin{center}
    $ s.t. u^T \dotprod (v_q-v_t) \geq 1-\epsilon_{qt}\geq 0 $
\end{center}

$\epsilon_{qt}$ is a very small non-negative value. The parameter $u$ can characterize the image sequence at the depth when $v_t$ has just started but $\nu_{t+1}$ has not yet proceeded which can be used as a feature descriptor of the depth image, which is a two-dimensional image at the same scale as the input image and contains information about the spatial and temporal variation of the entire motion. The information on the spatio-temporal variation of the whole action-forward process

The RankSVM method can be used for dynamic image generation. We used this algorithm to fuse 3D images into 2D images, and analogize the z-axis depth to the temporal dimension in the video.The parameter $u$ defines the order of the video frames $I_t$, indicating how the video frames evolve over time.

\subsection{Attention mechanism}
The CBAM(Convolutional Block Attention Module) algorithm is adopted to fuse spatial and temporal information for the attention module of our model. There are two separate submodules of CBAM, the Channel Attention Module (CAM) and the Spatial Attention Module (SAM). These two modules are responsible for extracting attention from the channel and space respectively, thus making the classification focus more on important parts. 
\begin{figure}[htp]
    \centering
    \includegraphics[width=8cm,height=6cm]{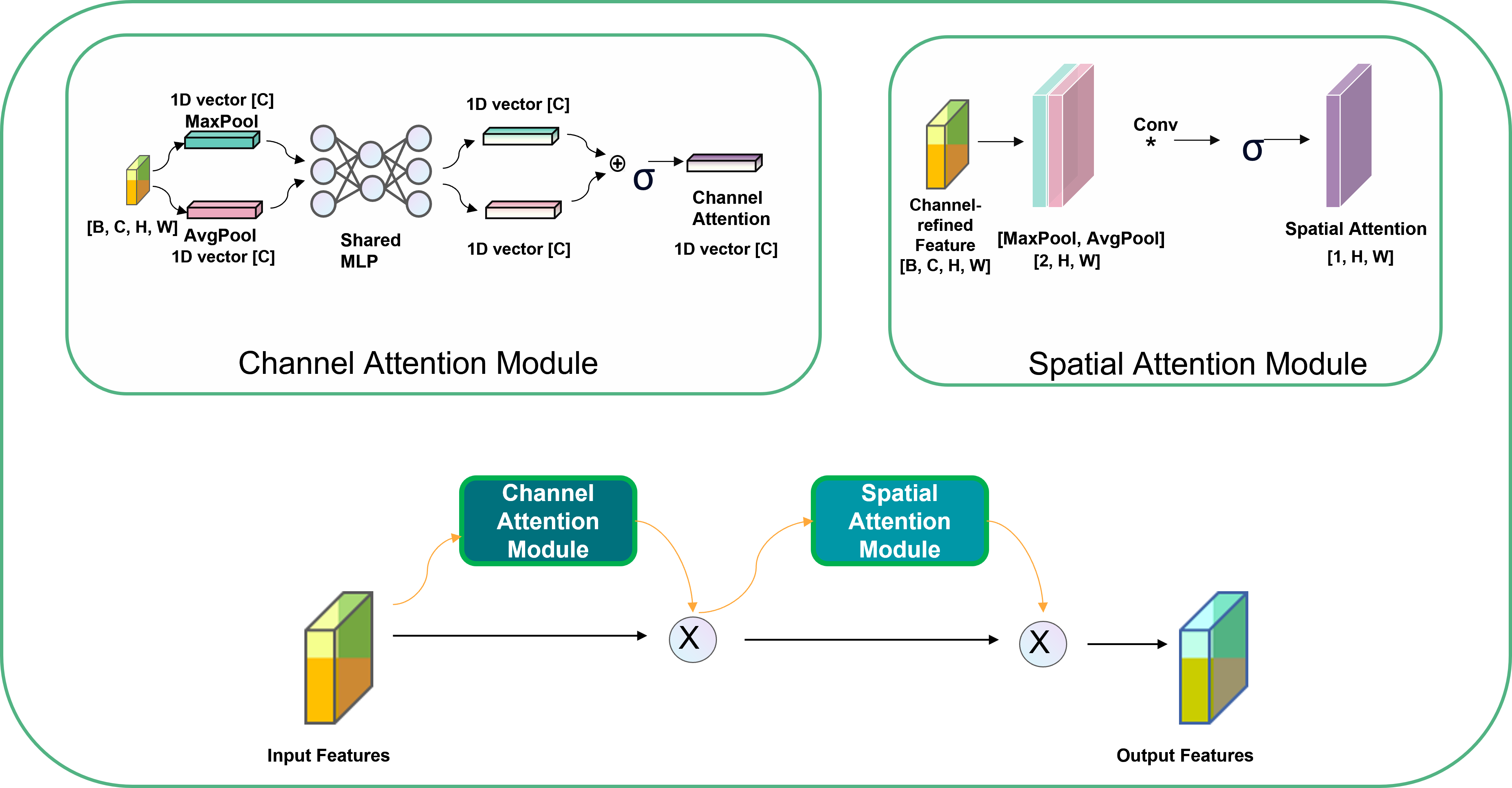}
    \caption{Attention mechanism}
    \label{fig:galaxy}
\end{figure}
The details of the CAM and SAM modules are shown in Fig.2 . In the channel attention module, the module first obtains the global spatial information of the input features through global average pooling and global max pooling operations. These channel operations compress the spatial dimensions of the feature maps to 1x1 while retaining the channel dimensions. It results in two 1D vectors, with each element representing a channel's average or maximum response. Next, these two vectors are mapped through a shared-weight fully connected layer (MLP) to obtain two new 1D vectors. These two vectors are combined through element-wise addition and a sigmoid activation function to generate the final channel attention weights. These weights will be used to adjust the channel response of the input feature maps, as shown in Fig.3. The formula definition is:
\begin{equation}
M_c(F) = \sigma(\text{MLP}(\text{AvgPool}(F)) + \text{MLP}(\text{MaxPool}(F))) = \sigma(W_1(W_0(F_{\text{avg}}) + W_1(W_0(F_{\text{max}}))))
\end{equation}

where $W_0$ and $W_1$ are learnable parameters of the MLP, and $\sigma$ denotes the sigmoid function. MLP is the Multilayer perceptron. The input size of this module is [B, C, H, W], where B stands for batch size, C stands for channels. The size of both one-dimensional vectors inside the Channel Attention Module is 256, which is the size of the input feature (C).

\begin{figure}[htp]
    \centering
    \includegraphics[width=8cm,height=1.6cm]{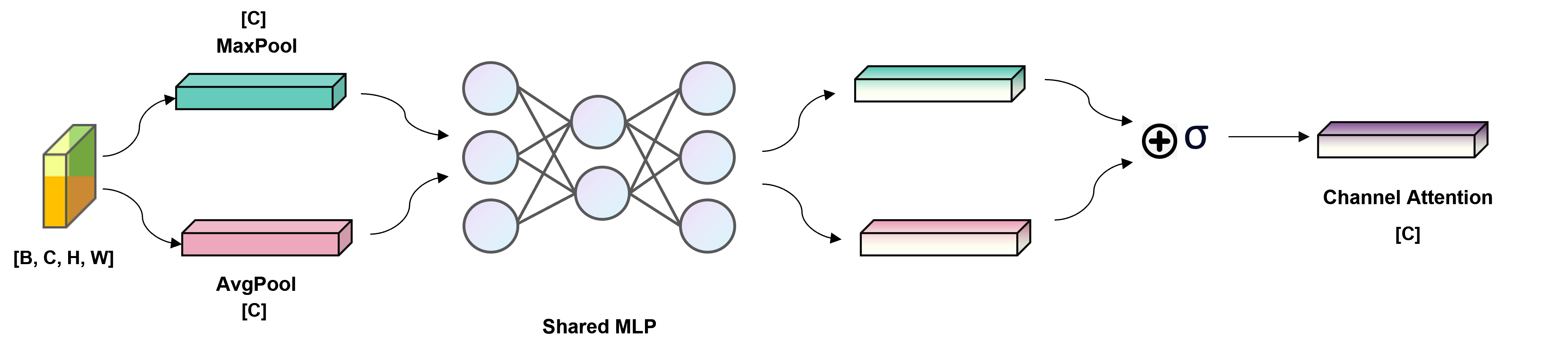}
    \caption{Channel attention module}
    \label{fig:galaxy}
\end{figure}

In the spatial attention module, the feature maps adjusted by the channel attention module are first subjected to global average pooling and global max pooling operations. They are performed along the channel dimension. The result is two 2D feature maps, with each element representing the average or maximum response at a spatial location. These two feature maps are combined by channel, where the result has two channel, passed through a 7*7 convolutional layer, and then processed with a sigmoid activation function to generate the final spatial attention weights and make this two-channel feature into one channel. The model structure is shown in Fig.4. These weights will be used to adjust the spatial response of the input feature maps. The formula definition is:
\begin{equation}
M_s\left(F\right)=\sigma\left(f\left[AvgPool\left(F\right);MaxPool\left(F\right)\right]\right)
\end{equation}
The size of the two 2D feature maps is (H, W).
\begin{figure}[htp]
    \centering
    \includegraphics[width=8.1cm,height=2.7cm]{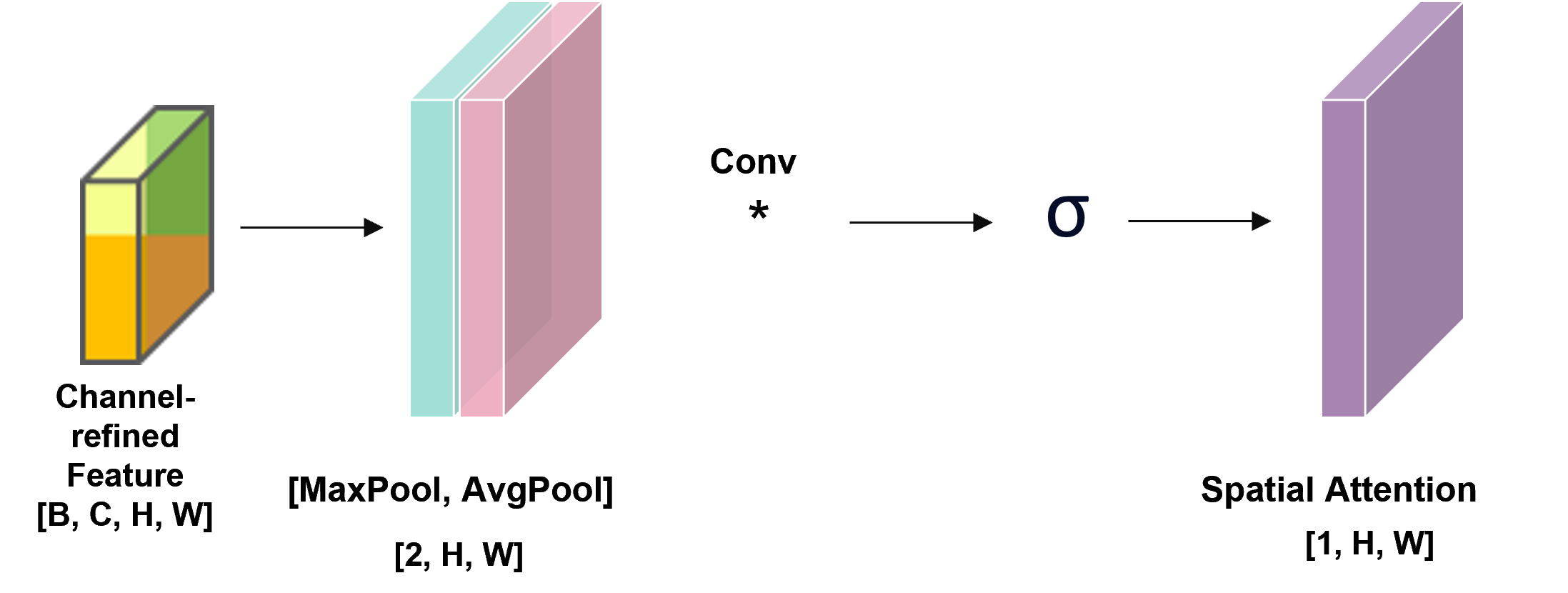}
    \caption{Spatial attention module}
    \label{fig:galaxy}
\end{figure}

\subsection{Classifier and loss function}
The fully connected layer is used as a classifier that integrates the information of all neurons, thereby obtaining the global information of all features for better classification. In the loss function, since the task of this study is a binary classification of Alzheimer's disease, binary cross-entropy is employed to better measure the difference between the model's prediction and the true value, making it easier to guide the model to adjust parameters in the direction of reducing prediction errors. For a 3D image V with label l and probability prediction p(l|V), the loss function is:
\begin{equation}
    loss(l,V)=-[l-log(p(l|V))+(1-l)\dotprod log(1-p(l|V))]
\end{equation} where l=0 indicates a normal sample (NC) and l=1 indicates an Alzheimer's disease sample (AD).

\section{Experiment result}
\subsection{Data Set}
We used the publicly available dataset from the Alzheimer's Disease Neuroimaging Initiative (ADNI) to testify the model. The brain MRI image size was 110*110*110. And the number of channels was 3. Since a subject may have multiple MRI scans in the database, we used the scan of each subject to avoid data leakage. The total number of data samples was 150, containing 75 NC samples and 75 AD samples. \par
This study implemented the model in the Ubuntu operating system, Intel i7-8700K 4core, 3.7GHz CPU, 64GB memory, and NVIDIA GTX 3090 platform. The programming language was Python, and the deep learning framework was PyTorch. The batch size used for our model was 16. The batch size of the baseline models was 8, the maximum batch size of the 3D. We used the Adam optimizer. The learning rate was 0.0001. We train for 150 epochs. To evaluate the performance of our model, we used accuracy (Acc), the area under the curve of Receiver Operating Characteristics (ROC), F1 score (F1), Precision, Recall, and Average Precision (AP) as our evaluation metrics.

\subsection{Data preprocessing}
The selected MRI images were preprocessed using T1 weighting and N3 bias field correction techniques. T1 weighted and skull-separated MRI images can be filtered from the ADNI database, and N3 bias field correction can be performed on these images using the functions provided in the simple ITK library. 

\begin{table}[]
    \caption{The comparison results for \\different CNN Backbone models}
    \label{CNNs}
    \centering
    \begin{tabular}{ c c c c c c c }
      \hline
      Model & ACC & AUC & F1 & Prec & Recall & AP \\ 
      \hline
      AlexNet & 0.87 & 0.84 & 0.86 & 0.89 & 0.83 & 0.82 \\
      VGG11 & 0.90 & 0.89 & 0.87 & 0.89 & 0.86 & 0.86 \\
      ResNet & 0.91 & 0.90 & 0.89 & 0.91 & 0.87 & 0.91\\ 
      \hline
    \end{tabular}
\end{table}

\subsection{Experiment results}
Experiments were conducted by replacing the CNN feature extractor in this study's model with other  CNN network structures, and the comparison results are shown in Table ~\ref{CNNs}. It is shown that the ResNet model used in this study outperformed other models. This is because the ResNet model adopts the residual block structure, which effectively avoids the vanishing gradient problem by adding skip connections in each convolutional layer.\par
It is shown that our model, shown as Post-Fusion-A in Table II, has the best performance on the evaluation indicator compared with 3D-VGG and Prefusion model. To further optimize the late fusion model, modifications were made by performing preliminary fusion on every ten slices of the three-dimensional MRI image and then conducting another fusion after passing through the 2D-CNN layer. However, the modified post-fusion model, shown as Post-Fusion-B in Table II, performed slightly worse than the post-fusion model. \par
In addition, experiments were conducted to replace the fully connected layer classifier with a binary SVM in the post-fusion model, with SVM parameters trained separately. However, the results show that the model, shown as Post-Fusion (SVM Classifier) in Table ~\ref{Model Results}, was less effective for classification compared to the fully connected layer approach.\par
By removing the attention module proposed in this study, the classification performance of the model, shown as Post-Fusion (Without CBAM) in Table ~\ref{Model Results}, was significantly reduced. This is because the attention mechanism in this study can adaptively learn channel and spatial attention on feature maps, accurately capturing the most discriminative information in the feature maps. Therefore, without this attention mechanism, the model losed this ability, leading to a decrease in classification performance.

\begin{table}[]
    \caption{Experiment results for different models}
    \label{Model Results}
    \centering
    \begin{tabular}{ c c c c c c c }
      \hline
      Model & ACC & AUC & F1 & Prec & Recall & AP \\ 
      \hline
        3D-VGG\cite{korolev2017residual} & 0.79 & 0.78 & 0.78 & 0.82 & 0.75 & 0.74\\
        Pre-Fusion\cite{xing2020dynamic} & 0.90 & 0.91 & 0.89 & 0.89 & 0.90 & 0.88\\
        Post-Fusion-A & 0.91 & 0.90 & 0.89 & 0.91 & 0.87 & 0.91\\
        Post-Fusion-B & 0.89 & 0.87 & 0.87 & 0.90 & 0.85 & 0.86\\
        Post-Fusion \\ (SVM Classifier) & 0.78 & 0.83 & 0.80 & 0.80 & 0.81 & 0.81\\
        Post-Fusion \\ (Without CBAM) & 0.82 & 0.83 & 0.83 & 0.84 & 0.83 & 0.76\\
        Post-Fusion \\ (MaxPooling) & 0.78 & 0.79 & 0.76 & 0.81 & 0.71 & 0.80\\

      \hline
    \end{tabular}
\end{table}

The model, shown as Post-Fusion (MaxPooling) in Table ~\ref{Model Results}, replaces temporal pooling with simple max pooling and had a significant decrease in the classification performance. This is because temporal pooling can effectively compress information in higher dimensions, thus improving the model's generalization ability. In contrast, max pooling simply takes the maximum values of each slice feature, failing to utilize the information in high-dimensional sequence data, resulting in a decrease in classification performance.

\begin{table}[]
    \caption{Training time for different models}
    \label{training time}
    \centering
    \begin{tabular}{ c c}
      \hline
      Model & Training time(s) \\ 
      \hline
        3D-VGG\cite{korolev2017residual} & 3438\\
        Post-Fusion-A & 1523\\
        Post-Fusion-B & 831\\
      \hline
    \end{tabular}
\end{table}
Table ~\ref{training time} shows the training time of different models. The parameter count of the 2D model was significantly lower than that of the 3D model. Concurrently, by performing early fusion on every ten slice images, the training time was substantially reduced in comparison to the original post-fusion model.

In our experiment, the loss function decline curve is shown in Fig.5 which is utilized to visualize the changes in the loss function values during the training process. Generally, it is expected that the value of the loss function decreases gradually as training progresses, indicating that the model is increasingly approaching the optimal solution. The following is an analysis of a loss function decline curve based on the Adam optimizer, assuming a training process of 150 epochs.

\begin{figure}[htp]
    \centering
    \includegraphics[width=8cm,height=6cm]{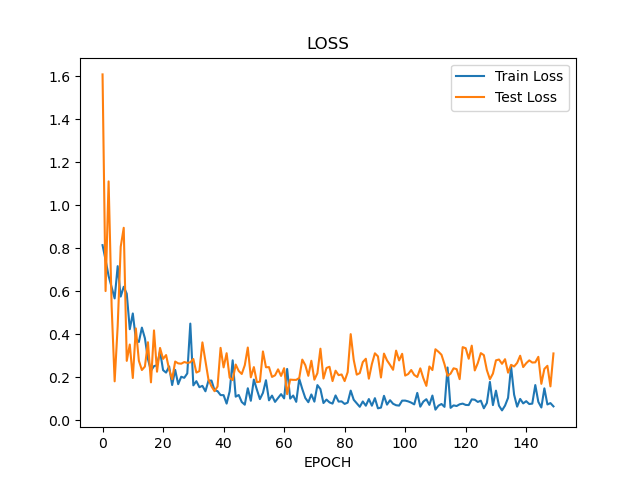}
    \caption{Loss function value versus epoch}
    \label{fig:galaxy}
\end{figure}

 In the first few epochs, the value of the loss function may exhibit substantial fluctuations. This was due to the model learning the training data in the initial stage, with the initialization of the weight parameters being relatively random, leading to instability in the loss function values. In the subsequent several dozen epochs, the value of the loss function usually declined rapidly. The Adam optimizer combined the characteristics of adaptive learning rates and momentum, allowing for the swift update of weight parameters in a more favorable direction during the early stages of training, thereby accelerating the decline of the loss function. As the value of the loss function gradually approached the optimal solution, the decline rate slowed down. In the 150-epoch training process, the loss function declined more rapidly in approximately the first half of the epochs, while gradually stabilizing in the latter half. In some epochs, the value of the loss function may exhibit minor jitter or oscillations. This may be due to factors such as noise in the training data, learning rate settings, or model parameter adjustments. The adaptive learning rate of the Adam optimizer can alleviate oscillations to some extent.

Overall, the loss function decline curve based on the Adam optimizer should exhibit a clear downward trend and tend to stabilize in the later stages of training. If the decline process of the loss function is not ideal, it may be necessary to consider adjusting the learning rate, increasing the amount of training data, or improving the network structure to enhance the model's performance.

\section{Conclusion}
Alzheimer's disease (AD) is a progressive neurodegenerative disorder that significantly impacts the cognitive and physical well-being of elderly individuals. Early detection of AD poses considerable challenges due to the subtle and intricate clinical manifestations during its initial phases. Computer-aided medical diagnosis utilizing image recognition techniques has emerged as a vital research area in this domain. In this study, we present an innovative Alzheimer's disease diagnostic model based on Convolutional Neural Networks (CNNs) in deep learning. The model employs a pre-trained ResNet network as its foundation, integrating dimensionality reduction techniques for 3D medical images and attention mechanisms. A substantial amount of real magnetic resonance imaging (MRI) data from AD patients was acquired from the publicly available ADNI database for model training and evaluation. The experimental outcomes reveal the exceptional performance of the proposed model in AD medical image recognition tasks, achieving a post-fusion F1 score of 0.89. The results suggest that the implemented 2D fusion algorithm substantially enhances the model's accuracy, while the incorporated attention mechanism effectively assigns importance to relevant regions in images, further boosting the model's diagnostic capabilities. In conclusion, this study introduces a cutting-edge Alzheimer's disease diagnostic model based on deep learning, which leverages a pre-trained ResNet network, dimensionality reduction methods for 3D medical images, and attention mechanisms. The proposed model demonstrates remarkable results in terms of accuracy and diagnostic efficacy, offering valuable contributions to the early detection and management of Alzheimer's disease
Our future work includes some preprocessing technologies (etc. image enhancement \cite{Tang2007ICIP} and noise reduction technologies\cite{Tang2010Speckle}\cite{Liu2011Speckle}) to improve it. We also plan to combine images and other information( e.g. gene information) in the network.

\bibliographystyle{unsrt}
\bibliography{sample-base}










\end{document}